\documentclass[aps,amsmath,amssymb,reprint,prl,superscriptaddress]{revtex4-2}
\usepackage{graphicx} 

\usepackage[english]{babel}

\newcommand{\be}{\begin{equation}}
\newcommand{\ee}{\end{equation}}
\newcommand{\bea}{\begin{eqnarray}}
\newcommand{\eea}{\end{eqnarray}}

\DeclareMathOperator{\Tr}{Tr}

\begin{document}

\title{Modelling time-irreversible avalanches}

\author{Andrea Baldassarri}
\affiliation{Istituto dei Sistemi Complessi - Consiglio Nazionale delle Ricerche, Piazzale A. Moro 2, I-00185, Rome, Italy}
\affiliation{Dipartimento di Fisica, Sapienza Università di Roma, Piazzale A. Moro 2, I-00185, Rome, Italy}
\author{Andrea Puglisi }
\affiliation{Istituto dei Sistemi Complessi - Consiglio Nazionale delle Ricerche, Piazzale A. Moro 2, I-00185, Rome, Italy}
\affiliation{Dipartimento di Fisica, Sapienza Università di Roma, Piazzale A. Moro 2, I-00185, Rome, Italy}
\affiliation{INFN, Sezione Roma2, Via della Ricerca Scientifica 1, I-00133, Rome, Italy}
\begin{abstract}
We investigate the problem of the time reversal symmetry of fluctuations, as witnessed by the average shape of avalanches. This quantity has been measured in a variety of systems, ranging from magnetic materials to earthquakes. Although an asymmetric shape is often observed, which is a signature of a non-equilibrium dynamics, there is no general theoretical control of this feature. In this paper, we propose a non equilibrium extension of a paradigmatic model for ``crackling-noise'', the so called ABBM model. Our model is strictly related to the Brownian Gyrator, which has been previously introduced in stochastic thermodynamics as the simplest model for thermal anisotropy, but it can also be framed in the context of rate-and-state models. It reproduces the phenomenology observed in experiments on granular friction, and allows for a systematic theoretical study of the asymmetry. We manage to correlate a measure of asymmetry, that can be easily computed in experiments, with the entropy production rates of the dynamics. 
\end{abstract}

\date{\today}

\maketitle

\paragraph{Introduction.} 

A historical example of exploitation of fluctuations to achieve crucial information about a physical system is the Einstein's theory of Brownian Motion~\cite{aspect2021einstein}. 
In the Langevin formulation, a fluctuating force appears as the coarse-grained effect of the collisions of the particle with the molecules of the liquid in thermal agitation.  In this specific case, the stationary  process is statistically time symmetric, since it describes an equilibrium dynamics. The study of fluctuations in  out-of-equilibrium dynamics, where time reversal invariance is broken, is an open and debated problem~\cite{sarracino2025nonequilibrium}.

\begin{figure}[t]
    \centering
\includegraphics[width=0.5\textwidth]{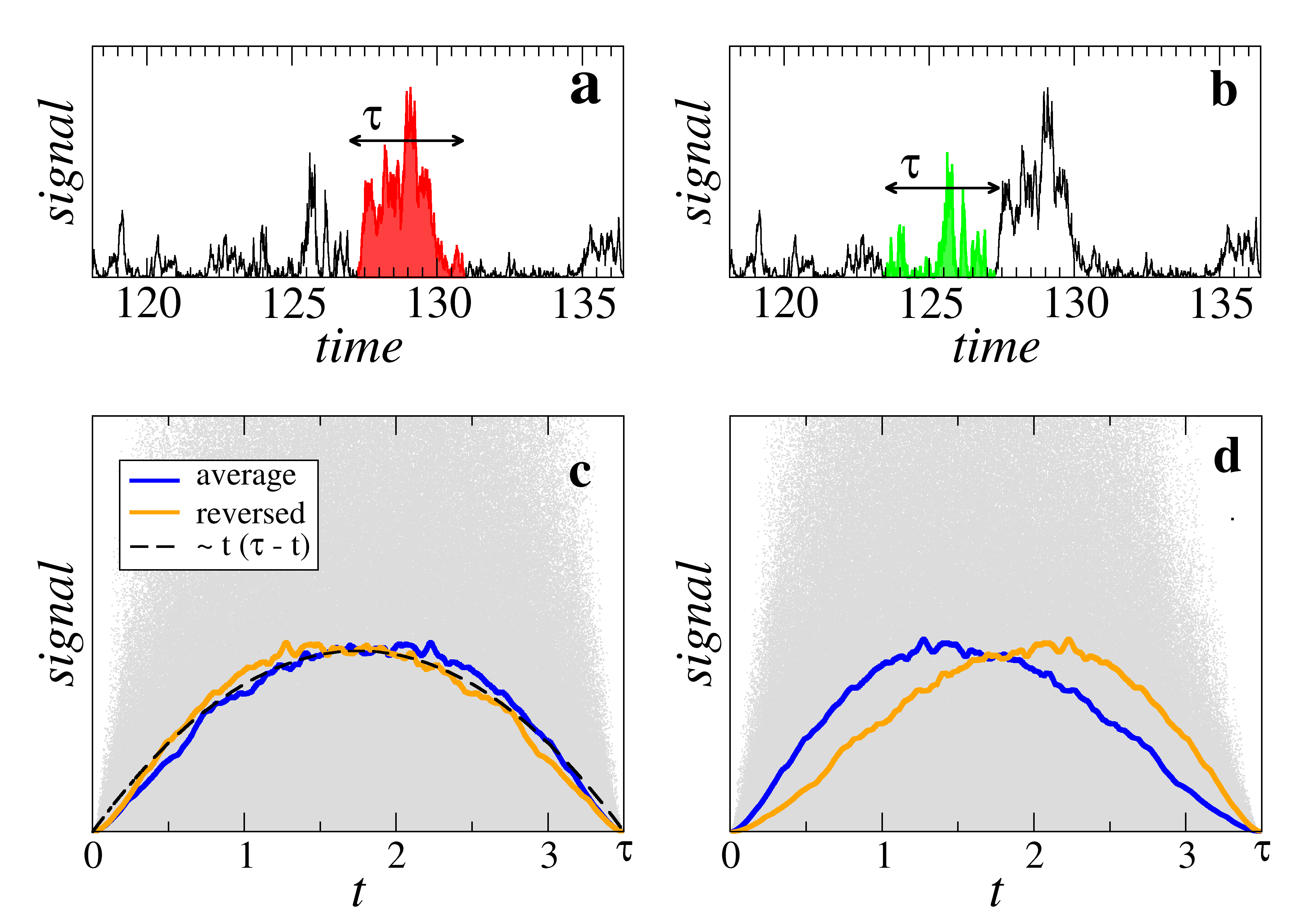} 
    \caption{Given a stochastic signal, an avalanche (red, panel a) and a bridge (green, panel b) are identified.  In panels c) and d) is depicted the definition of average shape: all the events (bridges in this case) of same duration $\tau$ are plotted in grey versus the internal time (which is zero at the start of the event): at each value of the internal time the average over the collected events gives the value of the average shape (blue line), resulting in a symmetric profile (example in panel c) or an asymmetric shape (example in panel d). The orange lines represents the reversed curve ($t\to \tau-t$) and it is plotted to manifest the asymmetry.  Data in figures are from arbitrary stochastic processes and used just for illustration.}
    \label{fig:avshapedef}
\end{figure}

There are many ways to characterize the fluctuations of a stochastic process $v(t)$. For phenomena with very intermittent and broadly distributed fluctuations (the so called ``crackling noisy'' systems~\cite{Sethna2001}), an analysis can be performed considering the signal as a sequence of ``avalanches" or ``burst", each one with its own duration and size (see Fig.~\ref{fig:avshapedef}a)~\cite{zapperi2022crackling}. Earthquakes obviously fall in this class, but a playground for these studies has been the Barkhausen noise, the very irregular short time dynamics of magnetization during hysteresis cycles of ferromagnetic materials~\cite{Bertotti1998}.  In this context, researchers proposed to measure the avalanche average shape $\langle v(t)
\rangle_\tau$, i.e. the time profile of the signal averaged over avalanches of the same duration $\tau$~\cite{Spasojevic1996,Mehta2002}. The idea is to capture and characterize directly the single fluctuation, inspecting its most visible feature, the time profile, also called the avalanche  {\em average shape} (see Fig.~\ref{fig:avshapedef}). 

Despite the appeal of such a measure, its theoretical study required a certain effort, even for simple processes~\cite{Baldassarri2003,Colaiori2008a,Papanikolaou2011,LeDoussal2012,Baldassarri2021}. Experimentally, the avalanche average shape has been measured in a variety of systems in material science~\cite{Chrzan1994,antonaglia14,ferrero16,Lagogianni2018,Sparks2018,sparks2017,toth_scaling_2023,Vu2020}, geophysics~\cite{Houston1998,bares_fluctuations_2014,Metha2006}, biology~\cite{Gallotti2018,Wang2013,gibiansky_earthquake-like_2013}, neuroscience~\cite{friedman_universal_2012,Roberts2014,Iyer2015,Wikstro2015,gleeson_temporal_2017,fontenele_criticality_2019,fontenele_criticality_2019,tian_theoretical_2022,alvankar_golpayegan_bistability_2023,capek_parabolic_2023,zaccariello_inhibitory_2025,deCandia2026}, astrophysics~\cite{Consolini2008,Sheikh2016}. 
The list of references collected here is not complete, but undergoes a physical criterion: the average shape measured in all these systems shows an {\em asymmetric profile}, witnessing the violation of the time reversal invariance.

In this letter we propose a model suitable for analytical study, featuring asymmetric average shapes  similar to those observed in experiments, for instance in granular friction~\cite{Baldassarri2019}, or more recently in luminosity bursts of white dwarves~\cite{dahmen2025}. 
The model, inspired to the Brownian Gyrator~\cite{filliger2007brownian}, popular in stochastic thermodynamics~\cite{Seifert2012}, is an extension of the ABBM model~\cite{Alessandro1990}, the paradigmatic model for crackling noise phenomena, but can also be regarded as a stochastic rate and state model~\cite{dieterich_time-dependent_1978,rice_stability_1983,rice_rate_2001}, because of its non Markovian character. We show that, in this case, the asymmetry
is associated to a breaking of the dynamical scaling, since it increases with avalanche durations $\tau$ up to a maximal value and then decreases. Interestingly, we observe that the skewness of the average shape is positively correlated with a measure of average entropy production, and in some cases can be expressed as a function of it.

\paragraph{Barkhausen noise and ABBM model.}~\label{sec:abbm}
One of the historical evidence of the existence of magnetic domains in a ferromagnetic materials was provided by Barkhausen in the 1919: The irregular noise produced by a change of an external field suggests an irregular motion of magnetic domain walls inside the material~\cite{Bertotti1998}. The statistical analysis of this crackling noise revealed that the magnetization changes through a sequence of bursts, with  quite diverse (broadly distributed) durations and amplitudes~\cite{Zapperi1997}. 
More recently~\cite{Spasojevic1996,Kuntz2000,Durin2002,Mehta2002},  the average shape of a single burst has been considered (see Fig.~\ref{fig:avshapedef}).
The simplest, yet successful model for the Barkhausen noise~\cite{Alessandro1990,Alessandro1990a} is the ABBM model that describes the velocity $v(t)$ of an ideal single magnetic domain wall. The model can be recast~\cite{Baldassarri2021} in terms of the stochastic process (previously introduced in finance, and known as the CIR process):
\be
\frac{dv}{dt} = k (c - v) + \sigma \sqrt{v} \,\eta(t), \label{CIR}
\ee
where $c$ is the rate of increase of the applied field and $k$ an effective stiffness of the domain wall.
A physical relevant point in Eq.~\eqref{CIR}, is the multiplicative nature of the noise term that comes from a spatially quenched disorder (at odds with thermal noise in Langevin theory of Brownian motion) and therefore disappears when $v=0$.  Eq.~\eqref{CIR} can be solved exactly~\cite{Feller1951} and it reproduces many statistical properties of the Barkhausen noise. Its average pulse shape is exactly computed~\cite{Papanikolaou2011,LeDoussal2012,Baldassarri2021}, and it results independent from the average wall velocity $c$, apart from a multiplicative factor (in Eq.~\eqref{bridge} we provide a more general expression that includes the ABBM case).
The shape in this case has a parabolic form $\propto t(\tau-t)$ in the limit of small avalanche duration $\tau<1/k$, and then flatten out for $\tau>1/k$. For any value of $\tau$ it obeys a $t\to \tau-t$ symmetry that descends directly from the invariance under time-reversal of the steady state of Eq.~\eqref{CIR}~\cite{baldassarri_bridge_2025}.
Such a symmetry is however in contrast with the pulse shape observed  for some ferromagnetic alloys~\cite{Durin2002,Mehta2002,spasojevic_barkhausen_2024}. 

\begin{figure}[t]
    \centering
\includegraphics[width=0.5\textwidth]{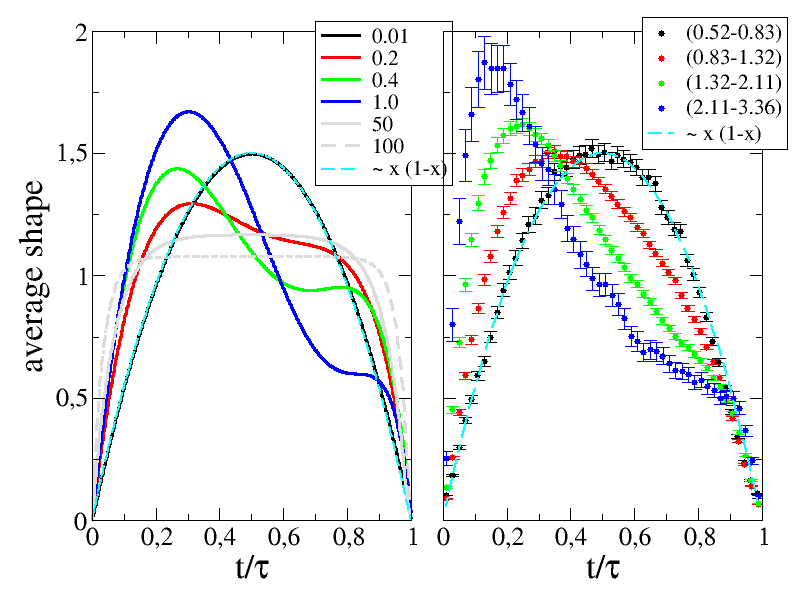} 
    \caption{Average shapes for model (left) and granular experiment~\cite{Baldassarri2019}  (right). At left, the average bridge shapes computed from model in Eq.~\eqref{BGeqs} are shown with increasing skewnesses  (parameters: $T_x = 0.1, T_y=100$, $\gamma_x=3, \gamma_y=0.5, u=1$). In the legend the corresponding durations. At right, average avalanche shapes from granular experiments with similar skewness (each curve is obtained averaging in a bracket of durations, as indicated in the legend) are quite similar with the scenario produced by the model.\label{fig:averagebridgeshapes}}
\end{figure}

\begin{figure}[h]
    \centering
        \includegraphics[width=0.5\textwidth]{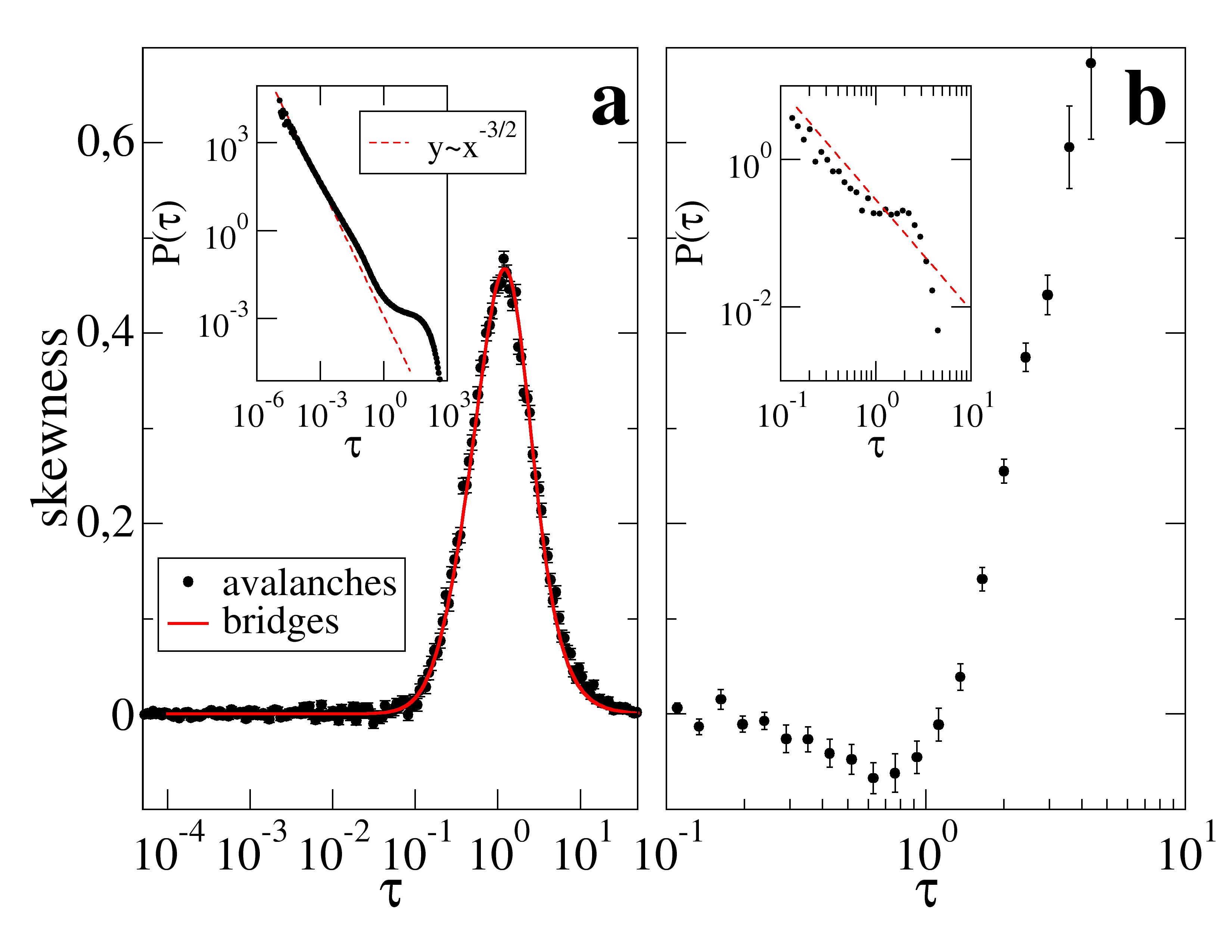}
    \caption{ Analysis of average shape skewness versus duration. In panel a) results from the model (same parameters as Fig.~\ref{fig:averagebridgeshapes}): red line is the theory for average bridge shapes, black points for average avalanche shapes (numerical). In panel b) the average skewness for average avalanche shapes versus their durations from granular experiment. In the insets the corresponding distributions of avalanche duration is shown for comparison. }
    \label{fig:comparisonthermalbgskewness}
\end{figure}

\paragraph{Asymmetric pulse shapes in granular friction.} Asymmetric pulse shapes are also observed in a variety of other systems that produces ``crackling noise" reminiscent of Barkhausen noise. For instance in~\cite{Baldassarri2019}, the angular velocity of a plate forced to slowly slide on a bed of granular materials is experimentally studied. For very slow driving, the plate performs a highly irregular motion, alternating periods of apparent sticks, with rapid and very diverse slips. The statistics of the angular velocity of the plate as a function of time  shares many analogies with Barkhausen noise, with the addition of significant inertial effects~\cite{Baldassarri2006}. 
In Fig.~\ref{fig:averagebridgeshapes}b, we reproduce the characteristic shapes observed in these experiments. For small $\tau$ we recover the symmetric parabolic shape observed in the ABBM model.
However, for large enough $\tau$ the average pulse shape~\cite{Baldassarri2019} violates the symmetry ($t\to \tau-t$), together with a breakdown of the  scaling behaviour (see inset in Fig.~\ref{fig:comparisonthermalbgskewness}b).  This is different from the scenario proposed in~\cite{Laurson2013}, where numerical data from models of depinning transition of driven elastic interfaces in random media, as well as from crack propagation experiments, have been studied. For these systems, the authors proposed a simple, possibly asymmetric, scaling function, whose fitted parameters would identify the universality class of the dynamics. 
A different approach was proposed earlier, in~\cite{Zapperi2005,Durin2007}, by reintroducing domain wall's inertia in the ABBM model. Numerical simulations demonstrated that the asymmetry observed in experiments can be reproduced (again apparently with the breakdown of the scaling behavior), but an analytical study seems hard to obtain.

In the following we propose a different extension to the ABBM model, that focuses on non equilibrium features, rather than inertia, but allows for an analytical study. This model has been recently proposed as one of the simplest theoretical test-ground for stochastic thermodynamics of non equilibrium systems and is often referred to as the Brownian Gyrator (BG).

\paragraph{Non-equilibrium ABBM extension.}
\label{sec:asymmetricpulse}

Let's first recall the connection between ABBM model and the two dimensional Ornestein-Uhlenbeck (OU) process, which is an equilibrium isotropic dynamics:
\be
\frac{d\vec r}{dt} = -\gamma \vec r   + \sqrt{2T} \vec \eta(t),\label{ouddim}
\ee
where $\vec r$ has coordinate $x,y$, while $\vec \eta$ is a vector of two independent white noises.
It is easy to change from the Cartesian to the polar coordinates $v=|\vec r|^2$, and $\theta$, using Ito's lemma:
\bea
\frac{d v}{dt} &=& 2( 2T  - \gamma v) +2\sqrt{2T v}\, \eta_1(t) \label{rhocir}\\
\frac{d\theta}{dt} &=& \sqrt{\frac{2T}{v}} \,\eta_2(t),
\eea
where $\eta_1$ and $\eta_2$ are two independent white noises. We recognize in the first equation the usual ABBM Eq.~\ref{CIR}, with parameters $c=2T/\gamma$, $k=2 \gamma$, and $\sigma = 2\sqrt{2T}$. Note that the stochastic process $v(t)$ is Markovian: Eq.~\ref{rhocir} is independent from $\theta$, as isotropy dictates.

The BG is a generalization of the OU-process~Eq.\eqref{ouddim} for anisotropic thermal fluctuations. In its original formulation~\cite{filliger2007brownian}, it can be considered as a model for the motion of an overdamped Brownian particle trapped in a 2-dimensional harmonic potential in contact with two different thermal baths~\cite{crisanti2012nonequilibrium,dotsenko2013two,baldassarri2020engineered}:
\be
\frac{d\vec r}{dt} = -\vec \nabla U(\vec r)  + \hat{D} \vec \eta(t),\label{BGeqs}
\ee
where the potential is parabolic  $U(\vec r)=\frac{1}{2} \gamma_x x^2 + \frac{1}{2} \gamma_y y^2 + u x y$, the matrix $\hat{D}$ has components $\hat{D}_{ij}=\delta_{ij} T_i$ where $i$ can be $x$ or $y$ and, crucially, $T_x$ and $T_y$ in general can be different. This model corresponds to a system with two coupled degrees of freedom in contact with two thermal baths. It is  simple to verify that the stationary distribution, although still Gaussian, differs from the Gibbs form $P_s(x,y) \propto \exp(-U/T)$, unless $T_x=T_y=T$. Moreover, the stationary state is characterized by a non zero probability current (solenoidal in the plane $x,y$) when $u\neq 0$ and $T_x\neq T_y$~\cite{dotsenko_two-temperature_2013}.

If we now consider the BG process in polar coordinates, we obtain a non equilibrium extension of the ABBM  model. The corresponding equations for $v=x^2+y^2$ and $\theta$ read:
\be
\begin{aligned}
\frac{dv}{dt} &=2\left[2\overline T- (\overline \gamma + 2\delta \gamma  \cos 2\theta )v\right]   + 2\sqrt{v} \eta_1 \\ 
\frac{d\theta}{dt} &= 2\delta \gamma \sin 2\theta - u \cos 2\theta +\frac {\delta T}{\sqrt{v}}\sin 2\theta + \frac 1{\sqrt{v}} \eta_2
\end{aligned}\label{abbm-extended}
\ee
where $\overline \gamma = (\gamma_x+\gamma_y)/2$, $\overline T = (T_x+T_y)/2$,  $\delta \gamma =\gamma_x-\gamma_y$, $\delta T = T_x-T_y$, and more importantly the two noises

\bea
\eta_1 &=& \cos\theta\sqrt{2T_x} \eta_x + \sin\theta\sqrt{2T_y}\eta_y \nonumber\\
\eta_2 &=& \sin\theta \sqrt{2T_x} \eta_x -\cos\theta\sqrt{2T_y}\eta_y \nonumber
\eea
are not independent, unless $T_x=T_y$, since
$\langle \eta_1(t) \eta_2(t') \rangle = (T_x - T_y) \sin 2\theta\, \delta(t-t')$. 

\paragraph{Average shapes.}
The computation of the average avalanche shape of the anisotropic process involves the use of its absorbing propagator (see End Matter, EM). It turns out that the computation of a similar quantity, called the average bridge shape of the process, can be performed. In this context the bridge can be defined as the snippet of dynamics trajectory between two zeros, not necessarily two consecutive ones (see Fig.~\ref{fig:avshapedef}b). These bridges can be considered as a sequence of avalanches with total duration $\tau$. For the ABBM process it has been shown~\cite{Baldassarri2019} that avalanche and bridge share the same average shape, apart for a proportionality constant. In principle it is not clear how much this equivalence is general for other diffusion processes. In the models considered in this letter, we prove it below for the equilibrium case $T_x=T_y=T$ and we numerically checked that it is always verified. 

First, let us proceed with the computation of the average bridge shape of duration $\tau$ in the equilibrium case ($T_x=T_y$). Its expression reads:
\be
\langle v(t)\rangle_\tau = \sum_{i=1,d}
\frac{2T}{\lambda_i} \frac{\sinh \left(\lambda_i (\tau-t)\right) \sinh (\lambda_i t)}{\sinh(\lambda_i \tau) },
\label{bridge}
\ee
where $\lambda_i$ are the eigen-values of $A$, defined as the symmetric Hessian matrix of the potential $U$.
Note that this is an average shape for a non Markov process ($v$ in Eqs.~\ref{abbm-extended}  may depends on $\theta$ even for $T_x=T_y$). However, as in the isotropic process, the  shape is symmetric for $t\to \tau-t$, as it should, since the underlying dynamics are both in equilibrium.

For  the non equilibrium case,
it is also possible to compute the average bridge shape for $v=x^2+y^2$. The final expression is $\langle v(t)\rangle_\tau = \Tr{S_B(t,\tau)}$, where the trace is computed on the covariance of the Gaussian one time probability function of the $x,y$ bridge process, which can be expressed as~\cite{Chen2016,lucente2022inference}:
\be
S_B(t,\tau) = S(t)-S(t)G^T(\tau-t)S^{-1}(\tau)G(\tau-t)S(t), \label{bridgecov}
\ee
in terms of $S(t)$, the covariance matrix of the free $x,y$ process, and $G(t)$, the response function $G(t)=\exp(-A t)$ (see EM for the derivation). The explicit expression as a function of the parameters $\gamma_x, \gamma_y, u, T_x$, and $T_y$ is too cumbersome to be shown here.  However it can be easily plotted for an arbitrary choice of the parameters. (A simple derivation of Eq.~\ref{bridgecov} is reported in the EM).

Our numerical analysis shows that the normalized average bridge shape coincides with the normalized average avalanche shape (see EM), and it looks symmetric for small duration  $\tau$ (Fig.~\ref{fig:averagebridgeshapes}a). However for larger $\tau$, a time-reversal asymmetry appears, and further increasing $\tau$ the shape starts to flatten. The flattening, however, seems often to appear first on the right part of the shape, giving rise to a peculiar left bulged average shape. Note the qualitative similarity with the scenario observed in granular experiments (Fig.~\ref{fig:averagebridgeshapes}b). Remarkably, very similar shapes are observed in the analysis of luminosity bursts of a white dwarf~\cite{dahmen2025}.

\paragraph{Shape skewness analysis.}
In order to better inspect the asymmetry of the average shape, we consider its skewness~\cite{Zapperi2005}. More precisely we consider the skewness of the  distribution defined as
\[
p_\tau(t) = \frac{\langle v(t)\rangle_\tau }{\int_0^\tau\langle v(t)\rangle_\tau dt}
\]
with respect to the average $\bar t = \int_0^\tau t\, p_\tau(t) dt$: A symmetric shape would produce $\bar t=\tau/2$ and zero skewness, while a positive (negative) skewness would indicate a left (right) asymmetry of the shape. (It is easy to show that the skewness does not depend on $\tau$ if the average shape assumes a scaling form $\tau^\alpha f(t/\tau)$.)

In Fig.\ref{fig:comparisonthermalbgskewness}a we plot the shape skewness as a function of the event duration $\tau$: the computation for bridges (same of avalanches within the numerical error) shows that the skewness attains a maximal value at $\tau=\tau_{max}$, corresponding with the maximally bulged shape, and then decreases toward zero, when the average shape flattens out (similar behaviour observed in~\cite{Zapperi2005}). In the companion panel, Fig.~\ref{fig:comparisonthermalbgskewness}b the same quantity is measured on the data of the granular experiment~\cite{Baldassarri2019}.
A similar increase of skewness is observed, but since large events are exponentially less probable, the expected decrease of the skewness (due to the decay of correlations for very large times) is not recorded (experimental statistics is over about 12000 avalanches, while the model allows for millions of events). The similitude between model and experiments is even more striking when looking at the duration distribution in the insets of Fig.~\ref{fig:comparisonthermalbgskewness}.

We investigated the dependence of the maximal skewness observed as a function of the model parameters. We considered many different choices of temperatures and potential parameters. For each choice  of parameters  $\gamma_x,\gamma_y,u,T_x,T_y$ we computed the skewness as a function of duration, and we identify the duration $\tau_{max}$ where the skewness reach its maximum. The results are thousands of different values of maximal skewness, ranging from $0$ to about $0.5$.

\begin{figure}[h]
    \centering
    \includegraphics[width=0.5\textwidth]{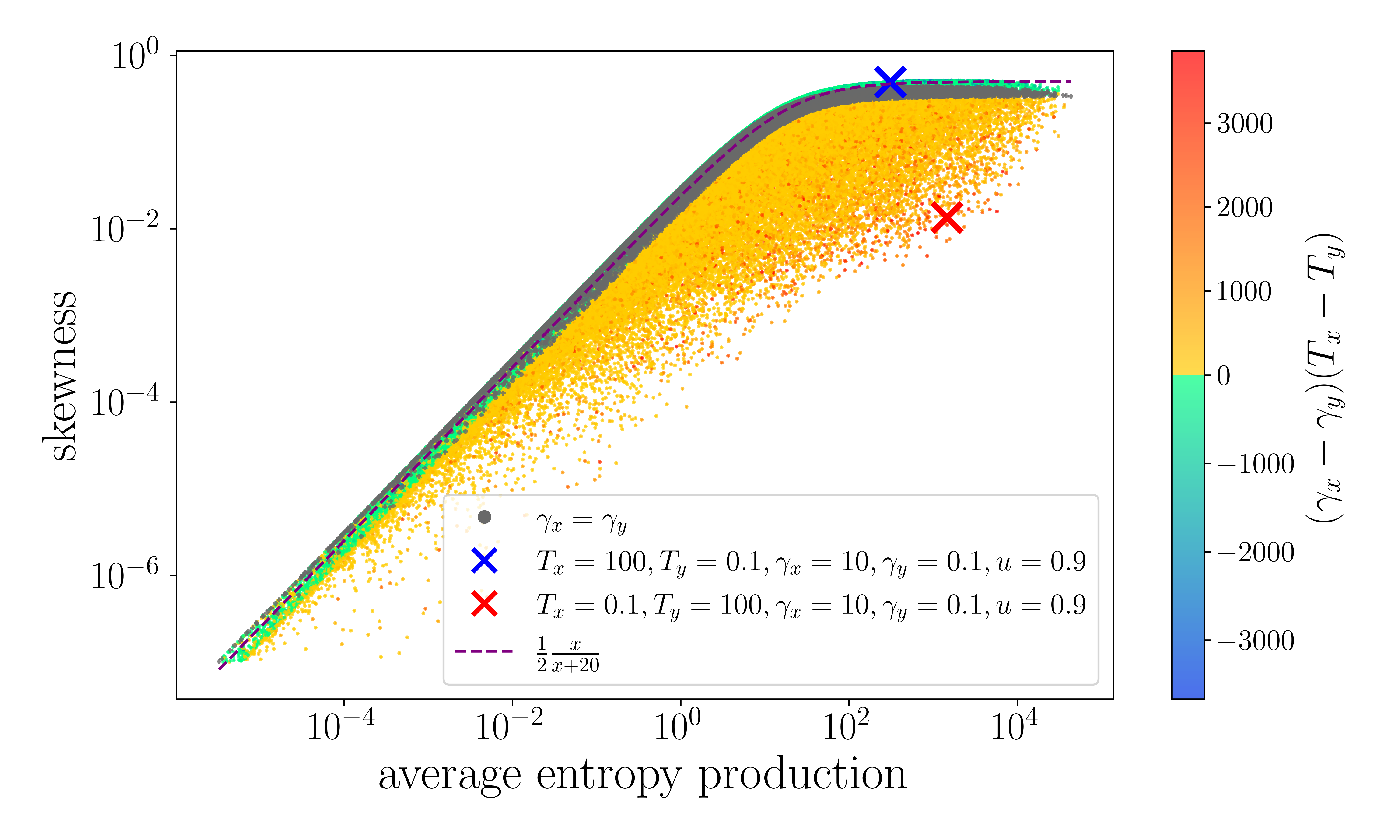}
    \caption{ Maximal skewness of the shape observed as a function of an estimation of the average entropy produced during the corresponding bridge. Points refers to more than 100000 different (random log-uniformly distributed) choices of parameters. The dashed line is a simple analytical expression describing qualitatively the case $(\gamma_x-\gamma_y)(T_x-T_y)\le 0$. Crosses represent two systems with the same drift parameters $\gamma_x, \gamma_y,$ and $u$, but temperatures swapped $T_x \leftrightarrow T_y$ (more details in End Material).}
    \label{Fig:scaling}
\end{figure}

Considering that the maximal skewness could be a measure of the irreversibility of the process, it is interesting to see how it correlates with the average entropy production of the most asymmetric bridge. We estimate this quantity as the average entropy production rate of the process multiplied by the duration $\tau_{max}$: $S_{prod} = \Phi \tau_{max}$, where the average production rate is~\cite{godreche_luck_2019,crisanti2012nonequilibrium}:
\be
\Phi = \frac{u^2}{\gamma_x+\gamma_y} \frac{\left(T_x-T_y\right)^2}{T_x T_y}.\label{entropyprodrate2}
\ee
The result is shown in Fig.~\ref{Fig:scaling} as a scatter plot: it shows a general positive correlation of skewness with average energy production. It also reveals an intriguing feature: The points relative to systems with parameters $\gamma_x=\gamma_y$ align along a thin line. Moreover, even when $\gamma_x\neq \gamma_y$, the points get very close to this ``collapsing" curve whenever $(\gamma_x-\gamma_y)(T_x-T_y)<0$
On the other hand, the most scattered cloud of points belong to the complementary class $(\gamma_x-\gamma_y)(T_x-Ty)>0$ (same number of points of  $(\gamma_x-\gamma_y)(T_x-Ty)<0$). At the moment we don't have an explanation for this observation.  In EM we show an emblematic extreme case, where two process sharing  the same average entropy production produce two different behaviours in terms of average bridge shape.

\paragraph{Discussion and conclusion.}\label{sec:conclusions}

In this paper we consider the problem of time irreversible dynamics witnessed by the asymmetry of average avalanche shapes. Previous studies in Barkhausen noise related the asymmetry of the average shape to a usually neglected effect of  magnetic wall inertia. This amounts to correct the ABBM model with a memory term that turns out to correspond to negative inertia~\cite{Zapperi2005,Durin2007}.
In friction studies, the introduction of memory effects has been considered notably within the so called rate-and-state framework~\cite{dieterich_time-dependent_1978,rice_stability_1983,rice_rate_2001}. There, the frictional shear strength depends on slip rate as well as on  the evolving properties of the contact population represented by a state variable with its own dynamical evolution.

For granular friction, previous attempts to compare experiments with rate-and-state models~\cite{bizzarri_earthquake_2021} or to introduce state variables in a stochastic context~\cite{leoni_friction_2011} are promising, but lack for a proper analytical treatment. In this work we propose an extension to the ABBM model that can be regarded as a stochastic rate-and-state friction model. In Eqs.~\ref{abbm-extended} the variable $\theta$ acts as an additional degree of freedom of the stochastic process, and breaks the Markovianity of the velocity dynamics, exactly as the state variable does in rate-and-state models. Rather than tackling the physical problem of microscopically modelling the granular contacts, we formulate the macroscopic model aiming for an analytical treatment. We leverage on the correspondence of the ABBM model with the polar representation of a two dimensional OU process, and consider one of the simplest out-of-equilibrium extension of this dynamic, the so called Brownian Gyrator. There, the system is kept out-of-equilibrium because of  anisotropic thermal baths. We notice that in fluidized granular setups, the dynamics of a massive intruder has been shown to be affected by hydrodynamics memory, as in molecular liquids, but with different effective temperature~\cite{Sarracino_2010}.

The result is in good qualitative agreement with experimental observations~\cite{Baldassarri2019}: average shape asymmetry appears when the scaling regime is broken, and depends on avalanche duration~\cite{Zapperi2005,Durin2007,Iyer2015} with bulged shapes of positive skewness (see Figs.~\ref{fig:averagebridgeshapes} and~\ref{fig:comparisonthermalbgskewness}). Similar shapes have been observed in very different systems~\cite{dahmen2025}, suggesting that a quite general mechanism is at play. 
We observe a notable positive correlation between the skewness of the average shape and the model’s entropy production, although no simple one-to-one relationship emerges.
This reveals that, despite its simplicity, the stochastic model displays rich phenomenology, motivating further theoretical work and encouraging comparison with other experiments.

\section*{Acknowledgments}

We thank Marco Baldovin, Dario Lucente and Massimiliano Viale for useful discussions. Alberto Petri and Andrea Gnoli for providing data from granular friction experiments.
Both Authors acknowledge funding from the Italian Ministero dell’Università e della Ricerca under the programme PRIN 2022 ("re-ranking of the final lists"), number 2022KWTEB7, cup B53C24006470006.

\bibliography{abbm-bg.bib}

@article{spasojevic_barkhausen_2024,
  title = {Barkhausen noise in disordered striplike ferromagnets: {Experiment} versus simulations},
  volume = {109},
  number = {2},
  journal = {Physical Review E},
  author = {Spasojevi\'c, Djordje and Marinkovi\'c, Milo\v{s} and Jovkovi\'c, Dragutin and Jani\'cevi\'c, Sanja and Laurson, Lasse and Djordjevi\'c, Antonije},
  month = feb,
  year = {2024},
  pages = {024110}
}

@article{zaccariello_inhibitory_2025,
	title = {Inhibitory neurons and the asymmetric shape of neuronal avalanches},
	volume = {111},
	issn = {2470-0045, 2470-0053},
	url = {https://link.aps.org/doi/10.1103/PhysRevE.111.024133},
	doi = {10.1103/PhysRevE.111.024133},
	number = {2},
	urldate = {2025-02-21},
	journal = {Physical Review E},
	author = {Zaccariello, Roberto and Herrmann, Hans J. and Sarracino, Alessandro and Zapperi, Stefano and De Arcangelis, Lucilla},
	month = feb,
	year = {2025},
	pages = {024133},
}

@article{deCandia2026,
  author={de Candia, Antonio  and Conte, Davide  and Golpayegan, Hanieh A.  and Scarpetta, Silvia },
  title={Symmetry breaking and avalanche shapes in modular neural networks},
  journal={Frontiers in Computational Neuroscience},
  volume={20},
  year={2026},
  url={https://www.frontiersin.org/journals/computational-neuroscience/articles/10.3389/fncom.2026.1744991},
  doi={10.3389/fncom.2026.1744991},
  ISSN={1662-5188}}

@article{gleeson_temporal_2017,
	title = {Temporal profiles of avalanches on networks},
	volume = {8},
	issn = {2041-1723},
	url = {https://www.nature.com/articles/s41467-017-01212-0},
	doi = {10.1038/s41467-017-01212-0},
	number = {1},
	urldate = {2026-05-08},
	journal = {Nature Communications},
	author = {Gleeson, James P. and Durrett, Rick},
	month = oct,
	year = {2017},
	pages = {1227},
}

@article{capek_parabolic_2023,
	title = {Parabolic avalanche scaling in the synchronization of cortical cell assemblies},
	volume = {14},
	issn = {2041-1723},
	url = {https://www.nature.com/articles/s41467-023-37976-x},
	doi = {10.1038/s41467-023-37976-x},
	number = {1},
	urldate = {2026-05-07},
	journal = {Nature Communications},
	author = {Capek, Elliott and Ribeiro, Tiago L. and Kells, Patrick and Srinivasan, Keshav and Miller, Stephanie R. and Geist, Elias and Victor, Mitchell and Vakili, Ali and Pajevic, Sinisa and Chialvo, Dante R. and Plenz, Dietmar},
	month = may,
	year = {2023},
	pages = {2555},
}

@article{Iyer2015,
    author = {Iyer, Kartik K. and Roberts, James A. and Hellström-Westas, Lena and Wikström, Sverre and Hansen Pupp, Ingrid and Ley, David and Vanhatalo, Sampsa and Breakspear, Michael},
    title = {Cortical burst dynamics predict clinical outcome early in extremely preterm infants},
    journal = {Brain},
    volume = {138},
    number = {8},
    pages = {2206-2218},
    year = {2015},
    month = {08},
    issn = {0006-8950},
    doi = {10.1093/brain/awv129},
    url = {https://doi.org/10.1093/brain/awv129},
    eprint = {https://academic.oup.com/brain/article-pdf/138/8/2206/13799903/awv129.pdf},
}

@incollection{aspect2021einstein,
  title={Einstein aujourd'hui},
  author={Aspect, Alain and Bouchet, Fran{\c{c}}ois and Brunet, {\'E}ric and Cohen-Tannoudji, Claude and Dalibard, Jean and Damour, Thibault and Darrigol, Olivier and Derrida, Bernard and Grangier, Philippe and Lalo{\"e}, Franck and others},
  booktitle={Einstein aujourd'hui},
  year={2021},
  publisher={EDP sciences}
}

@book{sarracino2025nonequilibrium,
  title={Nonequilibrium Statistical Mechanics: Basic concepts, models and applications},
  author={Sarracino, Alessandro and Puglisi, Andrea and Vulpiani, Angelo},
  year={2025},
  publisher={IOP Publishing}
}

@article{bizzarri_earthquake_2021,
	title = {Earthquake dynamics constrained from laboratory experiments: new insights from granular materials},
	volume = {64},
	issn = {2037-416X, 1593-5213},
	shorttitle = {Earthquake dynamics constrained from laboratory experiments},
	url = {https://www.annalsofgeophysics.eu/index.php/annals/article/view/8613},
	doi = {10.4401/ag-8613},
	
	number = {4},
	urldate = {2026-04-24},
	journal = {Annals of Geophysics},
	author = {Bizzarri, Andrea and Petri, Alberto and Baldassarri, Andrea},
	month = nov,
	year = {2021},
	pages = {SE441},
}

@article{rice_rate_2001,
	title = {Rate and state dependent friction and the stability of sliding between elastically deformable solids},
	volume = {49},
	issn = {00225096},
	number = {9},
	urldate = {2011-03-30},
	journal = {Journal of the Mechanics and Physics of Solids},
	author = {Rice, James R. and Lapusta, Nadia and Ranjith, K.},
	month = sep,
	year = {2001},
	pages = {1865--1898}
}

@article{rice_stability_1983,
	title = {Stability of {Steady} {Frictional} {Slipping}},
	volume = {50},
	issn = {0021-8936, 1528-9036},

	number = {2},
	journal = {Journal of Applied Mechanics},
	author = {Rice, J. R. and Ruina, A. L.},
	month = jun,
	year = {1983},
	pages = {343--349}
}

@article{dieterich_time-dependent_1978,
	title = {Time-dependent friction and the mechanics of stick-slip},
	volume = {116},
	issn = {1420-9136},
	number = {4},
	journal = {pure and applied geophysics},
	author = {Dieterich, James H.},
	month = jul,
	year = {1978},
	pages = {790--806}
}

@book{zapperi2022crackling,
  title={Crackling noise: statistical physics of avalanche phenomena},
  author={Zapperi, Stefano},
  year={2022},
  publisher={Oxford University Press}
}

@article{godreche_luck_2019,
  title = {Characterising the nonequilibrium stationary states of {Ornstein}-{Uhlenbeck} processes},
  volume = {52},
  issn = {1751-8113, 1751-8121},
  url = {https://iopscience.iop.org/article/10.1088/1751-8121/aaf190},
  doi = {10.1088/1751-8121/aaf190},
  
  number = {3},
  urldate = {2024-09-20},
  journal = {Journal of Physics A: Mathematical and Theoretical},
  author = {Godr\`eche, Claude and Luck, Jean-Marc},
  month = jan,
  year = {2019},
  pages = {035002},
}

@article{dotsenko_two-temperature_2013,
  title = {Two-temperature {Langevin} dynamics in a parabolic potential},
  volume = {87},
  copyright = {http://link.aps.org/licenses/aps-default-license},
  issn = {1539-3755, 1550-2376},
  url = {https://link.aps.org/doi/10.1103/PhysRevE.87.062130},
  doi = {10.1103/PhysRevE.87.062130},
  
  number = {6},
  urldate = {2024-07-25},
  journal = {Physical Review E},
  author = {Dotsenko, Victor and Macio\l{}ek, Anna and Vasilyev, Oleg and Oshanin, Gleb},
  month = jun,
  year = {2013},
  pages = {062130},
}

@article{Sarracino_2010,
doi = {10.1209/0295-5075/92/34001},
url = {https://doi.org/10.1209/0295-5075/92/34001},
year = {2010},
month = {nov},
publisher = {},
volume = {92},
number = {3},
pages = {34001},
author = {Sarracino, A. and Villamaina, D. and Gradenigo, G. and Puglisi, A.},
title = {Irreversible dynamics of a massive intruder in dense granular fluids},
journal = {Europhysics Letters}
}

@article{leoni_friction_2011,
  title = {Friction memory in the stick-slip of a sheared granular bed},
  volume = {357},
  issn = {00223093},
  url = {http://linkinghub.elsevier.com/retrieve/pii/S0022309310004965},
  doi = {10.1016/j.jnoncrysol.2010.07.046},
  number = {2},
  urldate = {2011-12-21},
  journal = {Journal of Non-Crystalline Solids},
  author = {Leoni, Fabio and Baldassarri, Andrea and Dalton, Fergal and Petri, Alberto and Pontuale, Giorgio and Zapperi, Stefano},
  month = jan,
  year = {2011},
  note = {Publisher: Elsevier B.V.},
  pages = {749--753},
}

@article{Mehta2002,
archivePrefix = {arXiv},
arxivId = {cond-mat/0112107},
author = {Mehta, Amit P. and Mills, Andrea C. and Dahmen, Karin A. and Sethna, James P.},
doi = {10.1103/PhysRevE.65.046139},
eprint = {0112107},
issn = {1063651X},
journal = {Physical Review E - Statistical Physics, Plasmas, Fluids, and Related Interdisciplinary Topics},
number = {4},
pages = {6},
primaryClass = {cond-mat},
title = {{Universal pulse shape scaling function and exponents: Critical test for avalanche models applied to Barkhausen noise}},
volume = {65},
year = {2002}
}

@article{Zapperi1997,
author = {Zapperi, Stefano and Cizeau, Pierre and Durin, Gianfranco},
journal = {Physical Review Letters},
number = {23},
pages = {4669--4672},
title = {{Dynamics of a ferromagnetic domain wall: Avalanches, depinning transition, and the Barkhausen effect}},
volume = {79},
year = {1997}
}

@article{Durin2002,
author = {Durin, G. and Zapperi, S.},
doi = {10.1016/S0304-8853(01)01077-0},
issn = {03048853},
journal = {Journal of Magnetism and Magnetic Materials},
pages = {1085--1088},
title = {{On the power spectrum of magnetization noise}},
volume = {242-245},
year = {2002}
}

@article{Alessandro1990a,
author = {Alessandro, Bruno and Beatrice, Cinzia and Bertotti, Giorgio and Montorsi, Arianna},
doi = {10.1063/1.346424},
issn = {00218979},
journal = {Journal of Applied Physics},
number = {6},
pages = {2908},
publisher = {{\$}abstract.copyright{\_}name.value},
title = {{Domain-wall dynamics and Barkhausen effect in metallic ferromagnetic materials. II. Experiments}},
url = {http://link.aip.org/link/doi/10.1063/1.346424/html},
volume = {68},
year = {1990}
}

@article{Alessandro1990,
author = {Alessandro, Bruno and Beatrice, Cinzia and Bertotti, Giorgio and Montorsi, Arianna},
doi = {10.1063/1.346423},
issn = {00218979},
journal = {Journal of Applied Physics},
number = {6},
pages = {2901},
publisher = {{\$}abstract.copyright{\_}name.value},
title = {{Domain-wall dynamics and Barkhausen effect in metallic ferromagnetic materials. I. Theory}},
url = {http://link.aip.org/link/doi/10.1063/1.346423/html},
volume = {68},
year = {1990}
}

@article{Laurson2013,
author = {Laurson, Lasse and Illa, Xavier and Santucci, St{\'{e}}phane and {Tore Tallakstad}, Ken and M{\aa}l{\o}y, Knut J{\o}rgen and Alava, Mikko J},
doi = {10.1038/ncomms3927},
issn = {2041-1723},
journal = {Nature communications},
month = {jan},
number = {0316},
pages = {2927},
pmid = {24352571},
title = {{Evolution of the average avalanche shape with the universality class.}},
url = {http://www.pubmedcentral.nih.gov/articlerender.fcgi?artid=3905775{\&}tool=pmcentrez{\&}rendertype=abstract},
volume = {4},
year = {2013}
}

@article{Baldassarri2021,
   author = {Andrea Baldassarri},
   doi = {10.1088/1742-5468/ac1404},
   issn = {1742-5468},
   issue = {8},
   journal = {Journal of Statistical Mechanics: Theory and Experiment},
   month = {8},
   pages = {083211},
   publisher = {IOP Publishing and SISSA},
   title = {Universal excursion and bridge shapes in ABBM/CIR/Bessel processes},
   volume = {2021},
   url = {https://iopscience.iop.org/article/10.1088/1742-5468/ac1404},
   year = {2021},
}

@article{Durin2007,
  title = {Signature of negative domain wall mass in soft magnetic materials},
  volume = {316},
  issn = {03048853},
  url = {http://linkinghub.elsevier.com/retrieve/pii/S0304885307004386},
  doi = {10.1016/j.jmmm.2007.03.213},
  number = {2},
  urldate = {2014-04-03},
  journal = {Journal of Magnetism and Magnetic Materials},
  author = {Durin, G. and Colaiori, F. and Castellano, C. and Zapperi, S.},
  month = sep,
  year = {2007},
  pages = {436--441},
}

@article{baldassarri_bridge_2025,
	title = {Bridge, {Reverse} {Bridge}, and {Their} {Control}},
	volume = {27},
	issn = {1099-4300},
	url = {https://www.mdpi.com/1099-4300/27/7/718},
	doi = {10.3390/e27070718},
	number = {7},
	urldate = {2025-11-03},
	journal = {Entropy},
	author = {Baldassarri, Andrea and Puglisi, Andrea},
	month = jul,
	year = {2025},
	pages = {718},
}

@book{blumenthal_excursions_1992,
	address = {Boston, MA},
	title = {Excursions of {Markov} {Processes}},
	isbn = {978-1-4684-9414-3},
	url = {http://link.springer.com/10.1007/978-1-4684-9412-9},
	doi = {10.1007/978-1-4684-9412-9},
	publisher = {Birkh\"auser Boston},
	author = {Blumenthal, Robert M.},
	year = {1992},
}

@article{Zapperi2005,
  title={Signature of effective mass in crackling-noise asymmetry},
  author={Zapperi, Stefano and Castellano, Claudio and Colaiori, Francesca and Durin, Gianfranco},
  journal={Nature Physics},
  volume={1},
  number={1},
  pages={46--49},
  year={2005},
  publisher={Nature Publishing Group}
}

@article{Baldassarri2006,
author = {A. Baldassarri and F. Dalton and A. Petri and S. Zapperi and G. Pontuale and L. Pietronero},
collaboration = {},
title = {Brownian Forces in Sheared Granular Matter},
publisher = {APS},
year = {2006},
journal = {Physical Review Letters},
volume = {96},
number = {11},
eid = {118002},
numpages = {4},
pages = {118002},
doi = {10.1103/PhysRevLett.96.118002}
}

@book{Bertotti1998,
author = {Bertotti, Giorgio},
title = {Hysteresis in Magnetism},
publisher = {Academic Press},
year = {1998}
}

@article{antonaglia14,
  title = {Bulk Metallic Glasses Deform via Slip Avalanches},
  author = {Antonaglia, James and Wright, Wendelin J. and Gu, Xiaojun and Byer, Rachel R. and Hufnagel, Todd C. and LeBlanc, Michael and Uhl, Jonathan T. and Dahmen, Karin A.},
  journal = {Phys. Rev. Lett.},
  volume = {112},
  issue = {15},
  pages = {155501},
  numpages = {5},
  year = {2014},
  doi = {10.1103/PhysRevLett.112.155501}
}

@article{Sparks2018,
title = "Shapes and velocity relaxation of dislocation avalanches in Au and Nb microcrystals",
journal = "Acta Materialia",
volume = "152",
pages = "86 - 95",
year = "2018",
issn = "1359-6454",
doi = "https://doi.org/10.1016/j.actamat.2018.04.007",
url = "http://www.sciencedirect.com/science/article/pii/S1359645418302805",
author = "G. Sparks and R. Maa{\ss}"
}

@article{crisanti2012nonequilibrium,
  title={Nonequilibrium and information: The role of cross correlations},
  author={Crisanti, Andrea and Puglisi, Andrea and Villamaina, Dario},
  journal={Physical Review E—Statistical, Nonlinear, and Soft Matter Physics},
  volume={85},
  number={6},
  pages={061127},
  year={2012},
  publisher={APS}
}

@article{baldassarri2020engineered,
  title={Engineered swift equilibration of a Brownian gyrator},
  author={Baldassarri, Andrea and Puglisi, Andrea and Sesta, Luca},
  journal={Physical Review E},
  volume={102},
  number={3},
  pages={030105},
  year={2020},
  publisher={APS}
}

@article{filliger2007brownian,
  title={Brownian gyrator: A minimal heat engine on the nanoscale},
  author={Filliger, Roger and Reimann, Peter},
  journal={Physical review letters},
  volume={99},
  number={23},
  pages={230602},
  year={2007},
  publisher={APS}
}

@article{dotsenko2013two,
  title={Two-temperature Langevin dynamics in a parabolic potential},
  author={Dotsenko, Victor and Macio{\l}ek, Anna and Vasilyev, Oleg and Oshanin, Gleb},
  journal={Physical Review E—Statistical, Nonlinear, and Soft Matter Physics},
  volume={87},
  number={6},
  pages={062130},
  year={2013},
  publisher={APS}
}

@article{Chen2016,
	title = {Stochastic {Bridges} of {Linear} {Systems}},
	volume = {61},
	issn = {00189286},
	doi = {10.1109/TAC.2015.2440567},
	number = {2},
	journal = {IEEE Transactions on Automatic Control},
	author = {Chen, Yongxin and Georgiou, Tryphon},
	year = {2016},
	note = {arXiv: 1407.3421},
	pages = {526--531},
}

@article{lucente2022inference,
  title={Inference of time irreversibility from incomplete information: Linear systems and its pitfalls},
  author={Lucente, D and Baldassarri, A and Puglisi, A and Vulpiani, A and Viale, M},
  journal={Physical Review Research},
  volume={4},
  number={4},
  pages={043103},
  year={2022},
  publisher={APS}
}

@article{gardiner1985handbook,
  title={Handbook of stochastic methods for physics, chemistry and the natural sciences},
  author={Gardiner, Crispin W},
  journal={Springer series in synergetics},
  year={1985},
  publisher={Springer Berlin Heidelberg}
}

@article{sparks2017,
title = {Shapes and velocity relaxation of dislocation avalanches in fcc and bcc  crystals},
author = {Sparks, G. and Sickle, J. and Dahmen, K.A. and Maa{\ss}, R.},
year = 2017,
journal = {arXiv:1705.06636}
}

@article{LeDoussal2012,
archivePrefix = {arXiv},
arxivId = {1104.2629},
author = {{Le Doussal}, P. and Wiese, K. J.},
doi = {10.1007/BF02379134},
eprint = {1104.2629},
issn = {00188220},
journal = {Europhys. Lett.},
pages = {46004},
title = {{Distribution of velocities in an avalanche}},
volume = {97},
year = {2012}
}

@article{Chrzan1994,
author = {Chrzan, D. C. and Mills, M. J.},
doi = {10.1103/PhysRevB.50.30},
issn = {01631829},
journal = {Physical Review B},
number = {1},
pages = {30--42},
title = {{Criticality in the plastic deformation of L12 intermetallic compounds}},
volume = {50},
year = {1994}
}

@article{Lagogianni2018,
author = {Lagogianni, Alexandra E and Liu, Chen and Martens, Kirsten and Samwer, Konrad},
doi = {10.1140/epjb/e2018-90051-7},
issn = {1434-6036},
journal = {The European Physical Journal B},
month = {Jun},
number = {6},
pages = {104},
title = {{Plastic avalanches in the so-called elastic regime of metallic glasses}},
url = {https://doi.org/10.1140/epjb/e2018-90051-7},
volume = {91},
year = {2018}
}

@article{Spasojevic1996,
author = {Spasojevi{\'{c}}, Djordje and Bukvi{\'{c}}, Srdjan and Milo{\v{s}}evi{\'{c}}, Sava and Stanley, H. Eugene},
doi = {10.1103/PhysRevE.54.2531},
issn = {1063651X},
journal = {Physical Review E - Statistical Physics, Plasmas, Fluids, and Related Interdisciplinary Topics},
number = {3},
pages = {2531--2546},
title = {{Barkhausen noise: Elementary signals, power laws, and scaling relations}},
volume = {54},
year = {1996}
}

@article{Wikstro2015,
author = {Wikstro, Sverre and Iyer, Kartik K and Roberts, James A and Hellstro, Lena and Pupp, Ingrid Hansen and Ley, David and Vanhatalo, Sampsa and Breakspear, Michael},
doi = {10.1093/awv147},
issn = {14602156},
journal = {Brain},
volume = {138},
pages = {2206--2218},
pmid = {26001723},
title = {{Cortical burst dynamics predict clinical outcome early in extremely preterm infants}},
year = {2015}
}

@article{Gallotti2018,
author = {Gallotti, Riccardo and Chialvo, Dante R},
journal = {Journal of The Royal Society Interface},
month = {Jun},
number = {143},
title = {{How ants move: individual and collective scaling properties}},
url = {http://rsif.royalsocietypublishing.org/content/15/143/20180223.abstract},
volume = {15},
year = {2018}
}

@article{Sheikh2016,
  title = {Avalanche Statistics Identify Intrinsic Stellar Processes near Criticality in KIC 8462852},
  author = {Sheikh, Mohammed A. and Weaver, Richard L. and Dahmen, Karin A.},
  journal = {Phys. Rev. Lett.},
  volume = {117},
  issue = {26},
  pages = {261101},
  numpages = {5},
  year = {2016},
  month = {Dec},
  publisher = {American Physical Society},
  doi = {10.1103/PhysRevLett.117.261101},
  url = {https://link.aps.org/doi/10.1103/PhysRevLett.117.261101}
}

@article{Consolini2008,
author = {Consolini, Giuseppe and {De Michelis}, Paola and Tozzi, Roberta},
doi = {10.1029/2008JA013074},
issn = {21699402},
journal = {Journal of Geophysical Research: Space Physics},
number = {8},
pages = {1--11},
title = {{On the Earth's magnetospheric dynamics: Nonequilibrium evolution and the fluctuation theorem}},
volume = {113},
year = {2008}
}

@article{Metha2006,
  title = {Universal mean moment rate profiles of earthquake ruptures},
  author = {Mehta, Amit P. and Dahmen, Karin A. and Ben-Zion, Yehuda},
  journal = {Phys. Rev. E},
  volume = {73},
  issue = {5},
  pages = {056104},
  numpages = {8},
  year = {2006},
  month = {May},
  publisher = {American Physical Society},
  doi = {10.1103/PhysRevE.73.056104},
  url = {http://link.aps.org/doi/10.1103/PhysRevE.73.056104}
}

@article{Baldassarri2003,
author = {Baldassarri, Andrea and Colaiori, Francesca and Castellano, Claudio},
journal = {Physical Review Letters},
number = {6},
pages = {60601},
publisher = {APS},
title = {{Average Shape of a Fluctuation: Universality in Excursions of Stochastic Processes}},
url = {http://link.aps.org/abstract/PRL/v90/e060601},
volume = {90},
year = {2003}
}

@article{Kuntz2000,
archivePrefix = {arXiv},
arxivId = {cond-mat/9911207},
author = {Kuntz, M. C. and Sethna, J. P.},
doi = {10.1103/PhysRevB.62.11699},
eprint = {9911207},
issn = {01631829},
journal = {Physical Review B - Condensed Matter and Materials Physics},
number = {17},
pages = {11699--11708},
primaryClass = {cond-mat},
title = {{Noise in disordered systems: The power spectrum and dynamic exponents in avalanche models}},
volume = {62},
year = {2000}
}

@article{Sethna2001,
author = {Sethna, James P and Dahmen, Karin A and Myers, Christopher R},
journal = {Nature},
number = {6825},
pages = {242--250},
title = {{Crackling noise}},
volume = {410},
year = {2001}
}

@article{Baldassarri2019,
author = {Baldassarri, A and Annunziata, M A and Gnoli, A and Pontuale, G and Petri, A},
journal = {Scientific Reports},
number = {1},
pages = {16962},
title = {{Breakdown of Scaling and Friction Weakening in Intermittent Granular Flow}},
volume = {9},
year = {2019}
}

@article{Vu2020,
author = {Vu, Chi-Cong and Weiss, J{\'{e}}r{\^{o}}me},
journal = {Physical Review Letters},
month = {sep},
number = {10},
pages = {105502},
pmid = {32955331},
title = {{Asymmetric Damage Avalanche Shape in Quasibrittle Materials and Subavalanche (Aftershock) Clusters}},
volume = {125},
year = {2020}
}

@article{gibiansky_earthquake-like_2013,
	title = {Earthquake-like dynamics in \textit{{Myxococcus} xanthus} social motility},
	volume = {110},
	issn = {0027-8424, 1091-6490},
	url = {https://pnas.org/doi/full/10.1073/pnas.1215089110},
	doi = {10.1073/pnas.1215089110},
	number = {6},
	urldate = {2026-05-07},
	journal = {Proceedings of the National Academy of Sciences},
	author = {Gibiansky, Maxsim L. and Hu, Wei and Dahmen, Karin A. and Shi, Wenyuan and Wong, Gerard C. L.},
	month = feb,
	year = {2013},
	pages = {2330--2335},
}

@article{toth_scaling_2023,
	title = {Scaling of {Average} {Avalanche} {Shapes} for {Acoustic} {Emission} during {Jerky} {Motion} of {Single} {Twin} {Boundary} in {Single}-{Crystalline} {Ni2MnGa}},
	volume = {16},
	issn = {1996-1944},
	url = {https://www.mdpi.com/1996-1944/16/5/2089},
	doi = {10.3390/ma16052089},
	number = {5},
	urldate = {2026-05-07},
	journal = {Materials},
	author = {Tóth, László Z. and Bronstein, Emil and Daróczi, Lajos and Shilo, Doron and Beke, Dezs\H{o} L.},
	month = mar,
	year = {2023},
	pages = {2089},
}

@article{Roberts2014,
author = {Roberts, James A and Iyer, Kartik K and Finnigan, Simon and Vanhatalo, Sampsa and Breakspear, Michael},
journal = {The Journal of Neuroscience},
month = {May},
number = {19},
pages = {6557 LP  -- 6572},
title = {{Scale-Free Bursting in Human Cortex following Hypoxia at Birth}},
volume = {34},
year = {2014}
}

@article{ferrero16,
  title = {Driving Rate Dependence of Avalanche Statistics and Shapes at the Yielding Transition},
  author = {Liu, Chen and Ferrero, Ezequiel E. and Puosi, Francesco and Barrat, Jean-Louis and Martens, Kirsten},
  journal = {Phys. Rev. Lett.},
  volume = {116},
  issue = {6},
  pages = {065501},
  numpages = {5},
  year = {2016},
  month = {Feb},
  publisher = {American Physical Society},
}

@article{bares_fluctuations_2014,
  title = {Fluctuations of {Global} {Energy} {Release} and {Crackling} in {Nominally} {Brittle} {Heterogeneous} {Fracture}},
  volume = {113},
  copyright = {http://link.aps.org/licenses/aps-default-license},
  issn = {0031-9007, 1079-7114},
  url = {https://link.aps.org/doi/10.1103/PhysRevLett.113.264301},
  doi = {10.1103/PhysRevLett.113.264301},
  
  number = {26},
  urldate = {2024-07-26},
  journal = {Physical Review Letters},
  author = {Bar\'es, J. and Hattali, M.L. and Dalmas, D. and Bonamy, D.},
  month = dec,
  year = {2014},
  pages = {264301},
}

@article{Wang2013,
  title = {Bursts of active transport in living cells},
  volume = {111},
  issn = {00319007},
  url = {https://link.aps.org/doi/10.1103/PhysRevLett.111.208102},
  doi = {10.1103/PhysRevLett.111.208102},
  number = {20},
  journal = {Physical Review Letters},
  author = {Wang, Bo and Kuo, James and Granick, Steve},
  month = nov,
  year = {2013},
  pmid = {24289710},
  note = {Publisher: American Physical Society},
  pages = {208102},
}

@article{friedman_universal_2012,
  title = {Universal critical dynamics in high resolution neuronal avalanche data},
  volume = {108},
  issn = {00319007},
  doi = {10.1103/PhysRevLett.108.208102},
  number = {20},
  journal = {Physical Review Letters},
  author = {Friedman, Nir and Ito, Shinya and Brinkman, Braden A.W. and Shimono, Masanori and Deville, R. E.Lee and Dahmen, Karin A. and Beggs, John M. and Butler, Thomas C.},
  year = {2012},
  pmid = {23003192},
  pages = {1--5},
}

@article{fontenele_criticality_2019,
  title = {Criticality between {Cortical} {States}},
  volume = {122},
  issn = {10797114},
  url = {https://doi.org/10.1103/PhysRevLett.122.208101},
  doi = {10.1103/PhysRevLett.122.208101},
  number = {20},
  journal = {Physical Review Letters},
  author = {Fontenele, Antonio J. and De Vasconcelos, Nivaldo A.P. and Feliciano, ThaÃ­s and Aguiar, Leandro A.A. and Soares-Cunha, Carina and Coimbra, BÃ¡rbara and Dalla Porta, Leonardo and Ribeiro, Sidarta and Rodrigues, Ana JoÃ£o and Sousa, Nuno and Carelli, Pedro V. and Copelli, Mauro},
  year = {2019},
  pmid = {31172737},
  note = {Publisher: American Physical Society},
  pages = {208101},
}

@article{tian_theoretical_2022,
  title = {Theoretical foundations of studying criticality in the brain},
  doi = {10.1162/netn_a_00269},
  journal = {Network Neuroscience},
  author = {Tian, Yang and Tan, Zeren and Hou, Hedong and Li, Guoqi and Cheng, Aohua and Qiu, Yike and Weng, Kangyu and Chen, Chun and Sun, Pei},
  year = {2022},
  pages = {1--38},
}

@article{alvankar_golpayegan_bistability_2023,
  title = {Bistability and criticality in the stochastic {Wilson}-{Cowan} model},
  volume = {107},
  issn = {2470-0045},
  url = {https://link.aps.org/doi/10.1103/PhysRevE.107.034404},
  doi = {10.1103/PhysRevE.107.034404},
  number = {3},
  journal = {Physical Review E},
  author = {Alvankar Golpayegan, Hanieh and de Candia, Antonio},
  month = mar,
  year = {2023},
  note = {Publisher: American Physical Society},
  pages = {034404},
}

@article{Houston1998,
  title = {Time functions of deep earthquakes from broadband and short-period stacks},
  volume = {103},
  issn = {0148-0227},
  doi = {10.1029/98JB02135},
  number = {B12},
  journal = {Journal of Geophysical Research},
  author = {Houston, Heidi and Benz, Harley M. and Vidale, John E.},
  year = {1998},
  pages = {29895},
}

@article{Papanikolaou2011,
author = {Papanikolaou, Stefanos and Bohn, Felipe and Sommer, Rubem Luis and Durin, Gianfranco and Zapperi, Stefano and Sethna, James P.},
doi = {10.1038/nphys1884},
issn = {1745-2473},
journal = {Nature Physics},
month = {jan},
number = {4},
pages = {316--320},
publisher = {Nature Publishing Group},
shorttitle = {Nat Phys},
title = {{Universality beyond power laws and the average avalanche shape}},
url = {http://dx.doi.org/10.1038/nphys1884},
volume = {7},
year = {2011}
}

@article{Feller1951,
author = {Feller, William},
doi = {10.2307/1969318},
isbn = {9780387952185},
issn = {0003486X},
journal = {Annals of mathematics},
number = {2},
pages = {173--182},
pmid = {21453821},
title = {{Two Singular Diffusion Problems}},
url = {http://www.jstor.org/stable/10.2307/1969318{\%}5Cnpapers3://publication/uuid/8EACF03F-CEBA-47DB-8D27-029BD918D24A},
volume = {54},
year = {1951}
}

@article{Seifert2012,
archivePrefix = {arXiv},
arxivId = {1205.4176},
author = {Seifert, Udo},
doi = {10.1088/0034-4885/75/12/126001},
eprint = {1205.4176},
issn = {00344885},
journal = {Reports on Progress in Physics},
number = {12},
title = {{Stochastic thermodynamics, fluctuation theorems and molecular machines}},
volume = {75},
year = {2012}
}

@article{Colaiori2008a,
author = {Colaiori, Francesca},
doi = {10.1080/00018730802420614},
issn = {0001-8732},
journal = {Advances in Physics},
month = {jul},
number = {4},
pages = {287--359},
title = {{Exactly solvable model of avalanches dynamics for Barkhausen crackling noise}},
url = {http://www.informaworld.com/openurl?genre=article{\&}doi=10.1080/00018730802420614{\&}magic=crossref{\%}7C{\%}7CD404A21C5BB053405B1A640AFFD44AE3},
volume = {57},
year = {2008}
}

@article{dahmen2025,
  title = {Interpreting luminosity bursts in a Kepler-measured ZZ Ceti using avalanche statistics},
  author = {Sickle, Jordan and Myers, Gabriel H. and Gausling, Kameron and Mullen, Ethan and Shah, Amartya and Kawaler, Steve and Dahmen, Karin A.},
  journal = {Phys. Rev. D},
  volume = {112},
  issue = {6},
  pages = {063059},
  numpages = {9},
  year = {2025},
  month = {Sep},
  publisher = {American Physical Society},
  doi = {10.1103/3kby-v1rp},
  url = {https://link.aps.org/doi/10.1103/3kby-v1rp}
}

\onecolumngrid

\section*{End Matter}
\paragraph{Excursion and bridge for a Markov stochastic process.}
In stochastic process theory the excursion is a one dimensional stochastic process that start from a value $v(0)=v_0$ and stay away from it for a time $\tau$, when it first comes back to $v(\tau)=v_0$~\cite{blumenthal_excursions_1992}. The usual definition of avalanche in statistical physics is a special excursion of a non-negative stochastic process (an ``activity'' of the system) where $v_0=0$ (or $v_0=\epsilon\to 0$ for practical reasons). It is possible to define an excursion from a generic time-homogeneous Markovian continuous stochastic process: in this case what is needed is the ``absorbing propagator'' $P_a(v_1,t|v_2)$ that is the probability to get a value $v(t)=v_1$ conditioned to the knowledge of a previous value $v(0)=v_2$ {\em and} to the condition $v(t')\neq v_0$ for $0<t' <t$~\cite{gardiner1985handbook}. For instance the one time probability for the excursion reads:
\be
P_E(v,t)=\frac{P_a(v,t|v_0)P_a(v_f,\tau-t|v)}{P_a(v_f,\tau|v_0)}. \label{excursion}
\ee
An analogous and often easier to compute process is the bridge, which shares the same definition of the excursion, but relaxing the condition during the flight, i.e. the stochastic trajectory going from $v(0)=v_0$ to $v(\tau)=v_f=v_0$ in a time $\tau$, whatever its value $v(t)$ for $t<\tau$ (one can also generalize with $v_f\neq v_0$). In the case of continuous Markov process, the one time distribution of the bridge $P_B(v,t)$ has an analogous expression as Eq.~\ref{excursion}, where now the usual ``free'' or unconditioned propagator $P(v_1,t|v_2)$ replaces of $P_a$. Note that both, excursion and bridge of a Markov process, are (time-non homogeneous) Markov processes, whose evolution are determined by a Fokker-Planck equation directly related to the Fokker-Planck equation of the original process (see~\cite{baldassarri_bridge_2025} for a more detailed introduction to the subject).
The average shape is simply defined by $\langle v(t)\rangle_{\tau} = \int v \,P_{\star}(v,t)\,dv$, where $P_{\star}$ is $P_E$ for the average shape of the excursion $\langle v(t)\rangle_{\tau}^{E}$, or $P_B$ for the average shape of the  bridge  $\langle v(t)\rangle_{\tau}^{B}$. In both cases, the time asymmetry of the average shape $\langle v(t)\rangle_{\tau}\neq \langle v(\tau-t)\rangle_{\tau}$ implies the time-irreversibility of the corresponding process $P_{\star}(v,t)\neq P_{\star}(v,\tau-t)$ (while the opposite may not be true~\cite{baldassarri_bridge_2025}).

\paragraph{Bridge for the Brownian Gyrator.}
Consider the Markov continuous stochastic process defined by:
\be
d\vec{x} = -A \vec{x} dt + B d\vec{W}, \label{Gauss}
\ee
where $\vec W$ is a (multi-dimensional) Wiener process, while $A$ is a drift matrix and $B$ a positive definite matrix defined, in two dimensions, by 
\[
\begin{array}{lr}
A = \begin{pmatrix}
\gamma_x & u \\
u & \gamma_y
\end{pmatrix},&
B = \begin{pmatrix}
\sqrt{2T_x}&0\\
0&\sqrt{2T_y}
\end{pmatrix}
\end{array}
\]
The formal solution of the stochastic Eq.~\ref{Gauss} is $\vec x(t) = G(t)\,\vec x(0) + \int_0^t G(t-t')\, B \,\vec W(t')dt'$ where  $G(t) = \exp(-A t)$ is the response function. Because of the linearity of the equation, the process is Gaussian, and the (free) propagator
is completely defined by mean and covariance matrix:
\bea
\langle \vec x(t)\rangle &=& G(t) \,\vec x(0)\\
S(t) &\equiv& \langle (\vec x(t) - \langle \vec x(t)\rangle)^T(\vec x(t) - \langle \vec x(t)\rangle)\rangle = \int_0^t G(t') Q G^T(t') dt' \label{bgcorr}
\eea
where $Q = B^T B$ (we give here the general relation valid for non symmetric matrices $A$ and $B$, and $\bullet^T$ is the transpose matrix operator). 
From Eq.~\ref{bgcorr} it can be shown that
\begin{equation}
S(t_1+t_2) = S(t_1)+G(t_1)S(t_2)G^T(t_1), \label{covariancerelation}
\end{equation}
which implies $\frac{d}{dt}S(t)= -A S(t)-S(t) A^T + Q$.  The stationary distribution, if it exists, is Gaussian with zero mean and covariance $\lim_{t\to\infty}S(t) = S_{st}$ that solves the Lyapunov equation $A S_{st}+S_{st}A^T = Q$. In this case, the non stationary covariance can also be expressed as $S(t) = S_{st} - G(t) S_{st}G^T(t)$.

Using its definition, it is easy to show that the bridge is again a (time non-homogeneous) Gaussian processes. Its one time distribution assumes, for $\vec x_0=\vec x_f=0$, a multi variate normal distribution with an inverse covariance matrix:
\be
S_B^{-1}(t,\tau)=S(t)^{-1}+G^T(\tau-t)S(\tau-t)^{-1}G(\tau-t).\label{invcovariance}
\ee
Using the Woodbury identity
$\left(W + U C V\right)^{-1} = W^{-1} - W^{-1}U \left(C^{-1}+VW^{-1}U\right)^{-1}V W^{-1}$, 
as well as the relation Eq.~\eqref{covariancerelation}, one can invert the matrix, getting~Eq.\ref{bridgecov}.

If we consider the polar coordinates, than the average bridge shape for the squared modulus $v=|\vec x|^2$ is nothing but the trace of the correlation matrix $S_B(t)$
\be
\langle v(t)\rangle^B_\tau = \langle |\vec x(t)|^2\rangle_\tau = \Tr S_B(t). \label{bridgeexpr}
\ee

\paragraph{Equivalence of avalanche and bridge shape.}

\begin{figure}[t!]
    \centering
    \includegraphics[width=0.5\textwidth]{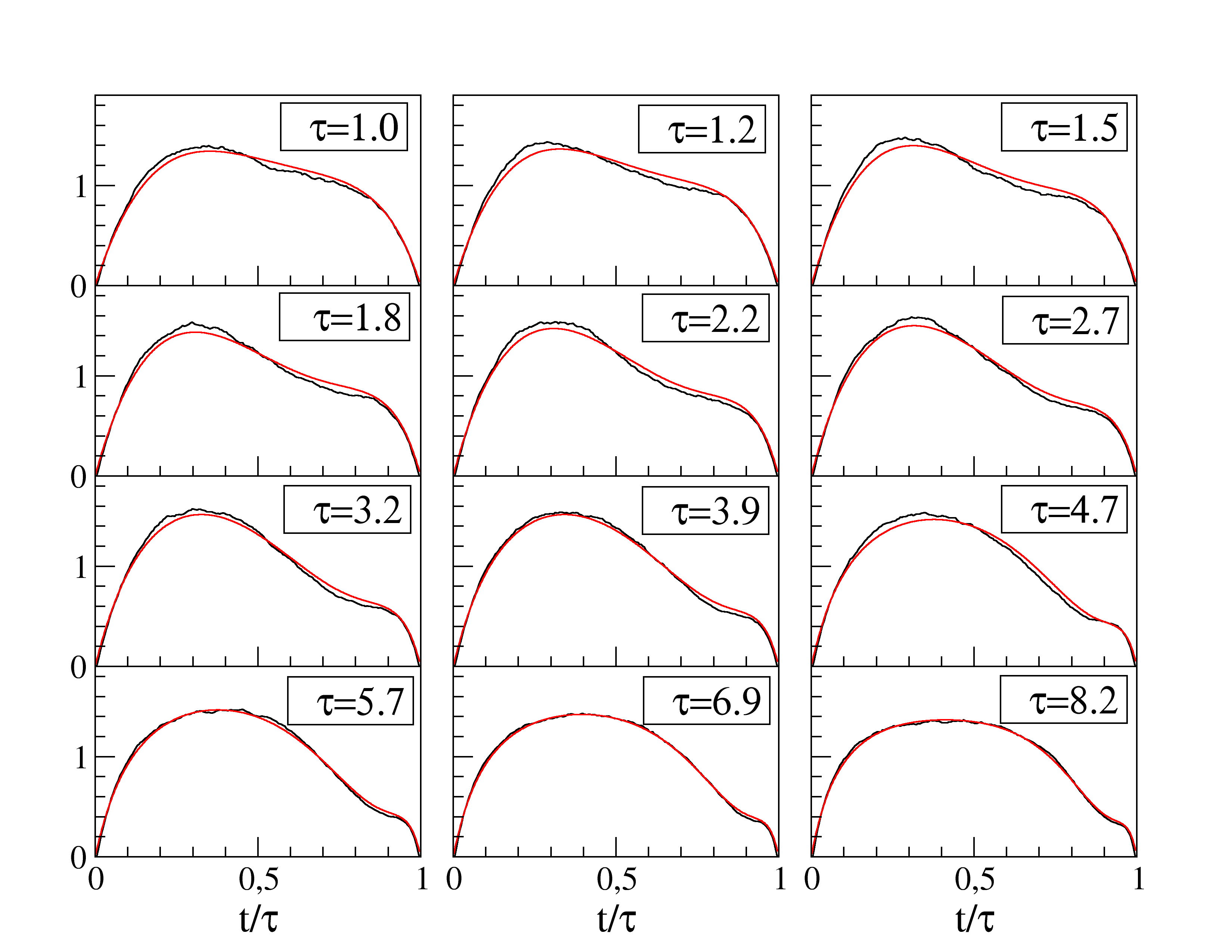} 
    \caption{Comparison of normalized average shapes for bridges (analytical) in red, and avalanche (numerical) in black. Parameters: $\gamma_x=\gamma_y=1$, $u=0.5$, $T_x=0.125$, $T_y=12.5$.  Only intermediate durations are shown: at smaller and larger durations, the shapes are  still indistinguishable and more symmetric, with respectively parabolic and flattening shapes.}
    \label{fig:compareshapes}
\end{figure}

In~\cite{baldassarri_bridge_2025}, an exact calculation for the ABBM/CIR process~Eq.\ref{CIR} shows that the bridge and avalanche (excursion) shapes are identical, apart from a proportionality constant. The model proposed in this paper is an out-of-equilibrium extension for the ABBM/CIR process, but we can not perform an analogous computation for the excursion, so we turn to numerical simulation in order to check the relation between shapes. In Fig.~\ref{fig:comparisonthermalbgskewness} we compare the skewness of the shapes, bridge (computed from Eq.~\ref{bridgeexpr}) and avalanche (from numerical integration of the Eq.~\ref{CIR}). In Fig.~\ref{fig:compareshapes} we show the direct comparison of shapes for several choices of the durations $\tau$ when $T_x\neq T_y$. We considered different parameters with similar results. The only care that has to be taken in the numerical simulations, is the value of the threshold $\epsilon$ for the definition of the excursion, that has to be small in order to get quantitatively precise results. See SI for more details on numerical implementation.

\begin{figure}[t!]
\centering
\includegraphics[width=0.45\textwidth]{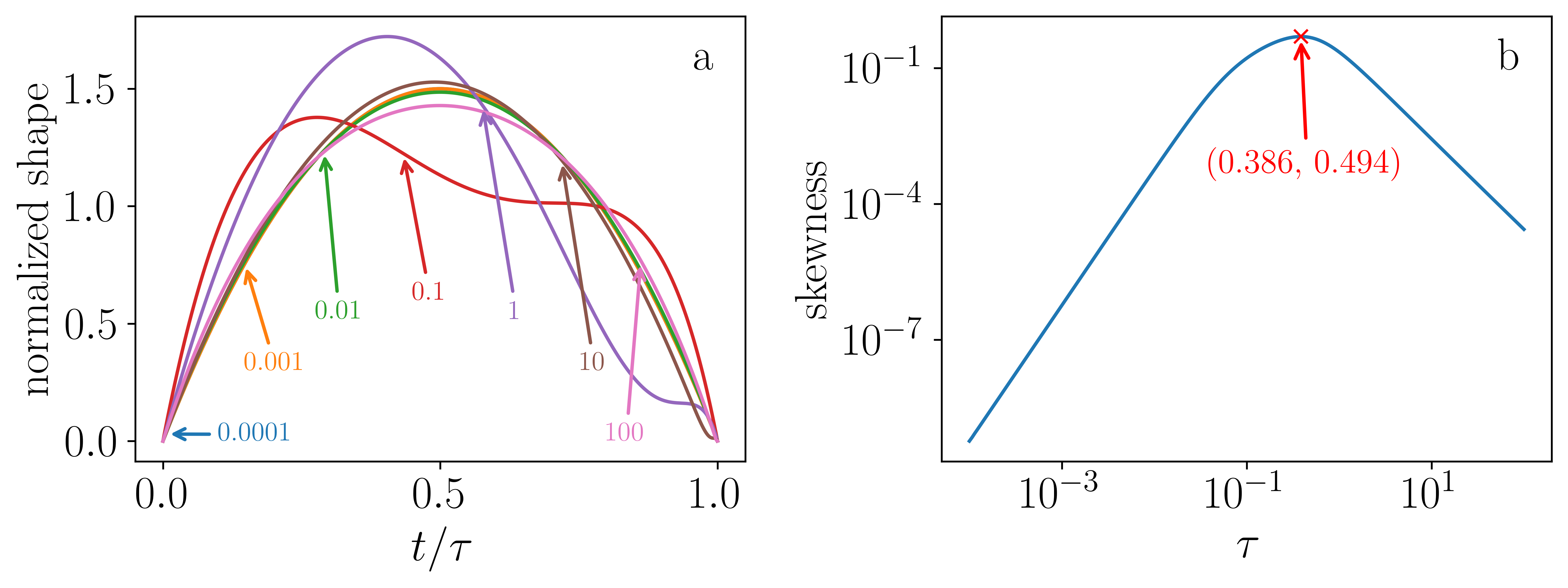}
\includegraphics[width=0.45\textwidth]{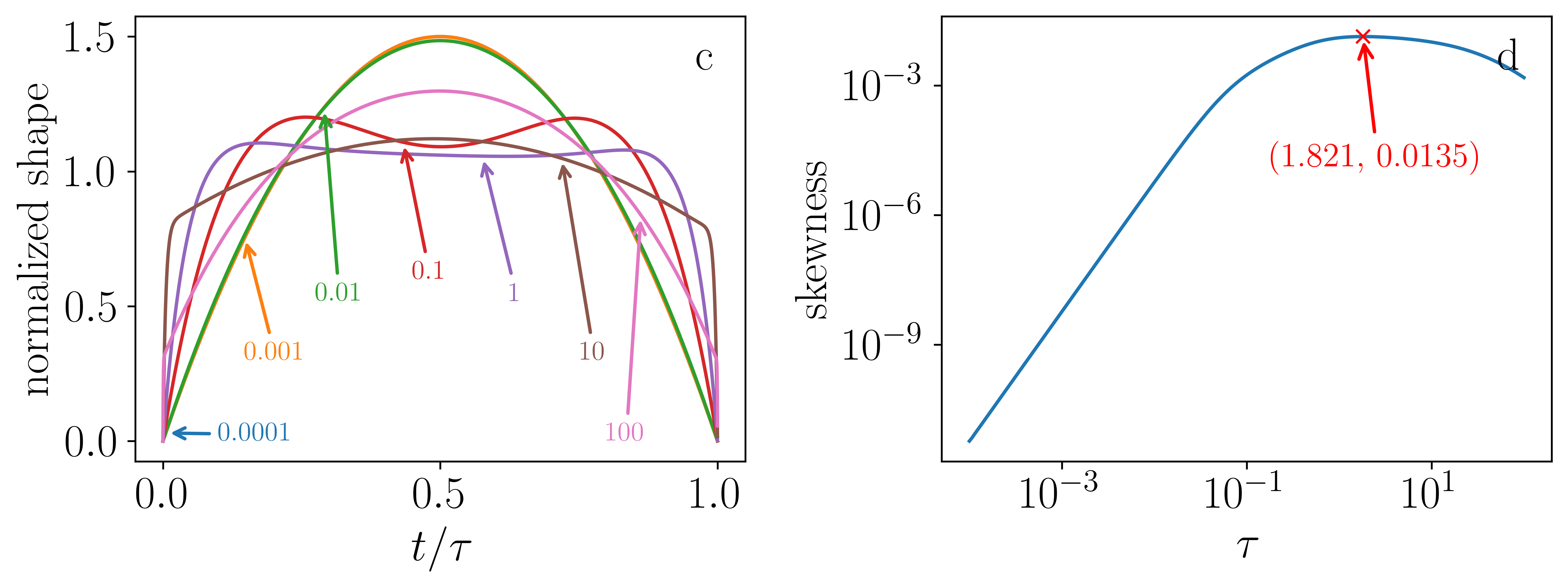} 
    \caption{Average shapes (a and c) and corresponding skewness (b and d) as a function of bridge duration $\tau$ (right) for two BG with same potential ($\gamma_x=10, \gamma_y=0.1, u=0.9$) but bath temperatures swapped. In panel a and b $T_x=0.01$ and $T_y=100$, while in panel c and d $T_x=100$ and $T_y=0.01$. Shapes are shown for ten different durations (see labels in panels a and d). In panels b and d, the point marked by a cross is the maximal skewness observed, whose values are reported in the label together with the corresponding duration.
    \label{fig:shape+skew}
    }
\end{figure}

\paragraph{An emblematic case of unexpected different skewnesses.} In Fig.~\ref{Fig:scaling} we show that the maximal skewness of the average shape is positively correlated with an estimation of the entropy production. However we noticed that, while the average entropy production rate of the process, Eq.~\ref{entropyprodrate2}, is symmetric with respect the swap $T_x\leftrightarrow T_y$, this is not the case for the maximal skewness of the average shape. In Fig.~\ref{fig:shape+skew} we show an extreme case: two systems sharing the same drift potential (same values of $\gamma_x$, $\gamma_y$ and $u$), but with temperature swapped $T_x\leftrightarrow T_y$.  The shapes drastically change, and the maximal skewness drops from about $0.5$ to about $0.01$. Note that also the duration of the maximal skewed shape $\tau_{max}$ changes, but it increases from about $0.4$ to $1.3$. Both systems share the same energy production rate (equal to $\approx 801.8$, according to Eq.~\ref{entropyprodrate2}). As a result, in the plot of Fig.~\ref{Fig:scaling}, the two systems results as two completely different points: the first (blue cross) lies very near the ``scaling line" (black points), with an abscissa about $x \approx 309$ and a skewness close to $0.5$, while the other (blue cross), at larger abscissa $x\approx 1460$ because of the larger $\tau_{max}$,has a much smaller skewness, about $0.01$.

\end{document}


\title{Supplemetary information for the letter {\em Modelling time-irreversible avalanches}}
\author{Andrea Baldassarri}
\affiliation{Istituto dei Sistemi Complessi - Consiglio Nazionale delle Ricerche, Piazzale A. Moro 2, I-00185, Rome, Italy}
\affiliation{Dipartimento di Fisica, Sapienza Università di Roma, Piazzale A. Moro 2, I-00185, Rome, Italy}
\author{Andrea Puglisi }
\affiliation{Istituto dei Sistemi Complessi - Consiglio Nazionale delle Ricerche, Piazzale A. Moro 2, I-00185, Rome, Italy}
\affiliation{Dipartimento di Fisica, Sapienza Università di Roma, Piazzale A. Moro 2, I-00185, Rome, Italy}
\affiliation{INFN, Sezione Roma2, Via della Ricerca Scientifica 1, I-00133, Rome, Italy}

\maketitle
\onecolumngrid 
\section{Numerical computation of average excursion shape and its skewness for the generalized ABBM model.}

In order to compute the average excursion shape for the generalized ABBM model, we proceeded with a simple numerical integration (Euler–Maruyama method) of the stochastic differential equation for the underlying two dimensional BG model:
\[
\frac{d\vec r}{dt} = -A \vec r  + \hat{D} \vec \eta(t)
\]
(see Eq.5 in the main text for notation).
We considered an excursion as the snippet of the trajectory of $v(t)=|\vec r|^2$ between two consecutive crossing of a small threshold value $\epsilon$. In order to obtain the precision needed to compare with our model, care has to be taken in the choice of a small enough $\epsilon$, as well as a small integration step $dt$. 

For the skewness, one should compute the average shape of a given duration (one has to choose a range of durations between $\tau$ and $\tau+\Delta \tau$, with $\delta \tau$ chosen not too small, in order to not compromise the number of excursion collected). We performed this task for several durations. 

However, we also considered a different approach. We measured the skewness of each excursion detected, collecting a large number of skewnesses as a function of $\tau$. Then we averaged the skewness of excursions of similar durations (binning the range of $\tau$ observed).

In principle this average is different from what computed in the model, since it represent the average skewness of the excursions of a given duration, while in the model we consider the skewness of the average shape of duration $\tau$. Since the skewness is not a linear function, the two quantities may differ. However, we checked that the numerical computation of the average skewness of excursions (which is a much simpler to compute) coincides with the skewness of the average shape on a set of durations spanning the range we considered for our analysis.
Both measures, coincide, within the numerical precision, with the skewness of the average bridge shapes computed from their analytical expression.

\section{Numerical analysis for skewness}
\begin{figure}[h]
    \centering
    \includegraphics[width=0.9\textwidth]{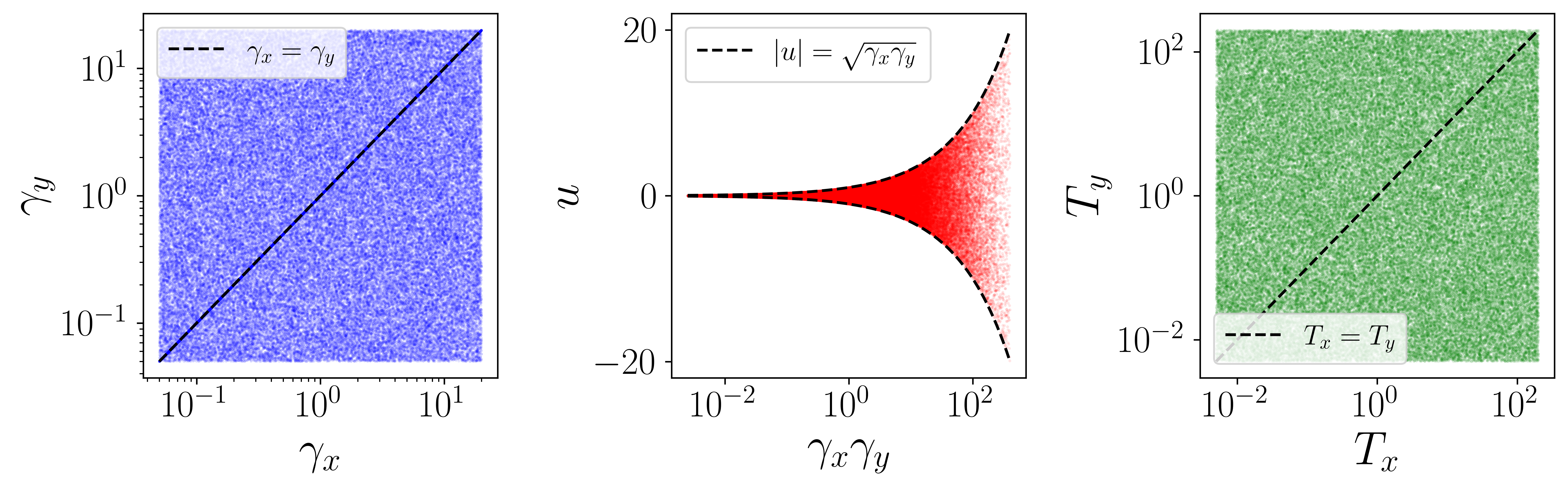} 
    \caption{Set of parameters considered for the study of maximal skewness of the thermal BG, for a total of  $100.000$ cases. Panel a) shows the possible choices of $\gamma_x$ and $\gamma_y$; panel b) shows the values considered for $u$, with $u^2<\gamma_x \gamma_y$ in order to have a stable potential $U$; panel c) shows the choices for $T_x$ and $T_y$. Equilibrium cases ($T_x=T_y$ or $u=0$) are excluded from the analysis.}
    \label{Fig:parameters}
\end{figure}

We spanned the parameter space of the model considering random parameters $\gamma_x, \gamma_y, u, T_x, T_y$, with uniform distribution in logarithmic scale. In Fig.~\ref{Fig:parameters} we show the actual values used, for a number of more than 100000 possibilities.
For each choice of parameters, we computed the average bridge shape of $v(t)$ as a function of the duration $\tau$.  The computation is performed with a simple python script, by menans of the SciPy linear algebra function scipy.linalg.solve\_continuous\_lyapunov() in order to compute the stationary covariance using the Lyaponov equation:
\[
S_{st} = Q A - A Q.
\]
Then, for a time $t$ and its symmetric value $\tau-t$, we computed the response function, using the scipy.linalg.expm() and the definition:
\[
G(t)=\exp (-A t).
\]
This allows to compute the covariance $S(t)= S_{st}  -  G(t) S_{st} G(\tau-t)$ (expression valid for symmetric drift matrix, for which $G=G^T$), and from that we obtained the covariance of the bridge distribution:
\[
S_B(t,\tau) = S(t)-S(t)G(\tau-t)S^{-1}(\tau)G(\tau-t)S(t),
\]
where we also needed $S^{-1}(\tau)$ computed via the numpy function numpy.inv.
The procedure is quite simple, the only care is to be taken for the computation of the response function, which can raise some underflow/overflow warning outside the range of parameter we considered.

A simple Newton-like algorithm allowed to identify the duration $\tau_{max}$ giving the maximal skewed shape. The success of this procedure is due to the simple functional behavior observed of skewness versus $\tau$, which shows a single maximum. Care has to be taken for sytems with $T_x \approx T_y$ or $u$ small, since in this case the skewness remains very low and its numerical computation can fluctuate. In this case the noise can make difficult the determination of the maximum. However, we managed to obtain good estimation even for very small skewness ($<10^{-6}$).

\bibliography{abbm-bg.bib}